\let\ifarxiv=\iftrue     
{}
{}

\newif\ifpublic\publictrue

\ifarxiv

\documentclass[12pt,a4paper]{article}

\usepackage{amsmath,amssymb}
\ifpublic\else\usepackage{showkeys}\fi
\usepackage{graphicx}
\usepackage[bookmarks=true,hyperfigures=true]{hyperref}
\usepackage[usenames, dvipsnames]{color}
\usepackage[nosort]{cite}
\usepackage[bulletsep]{collref}

\def\showkeysrefformat#1{{\normalfont\tiny\ttfamily#1}}
\makeatletter
\def\SK@@ref#1>#2\SK@{%
 {\@inlabelfalse\leavevmode\vbox to\z@{%
 \vss\SK@refcolor\rlap{\vrule\raise .75em%
  \hbox{\showkeysrefformat{#2}}}}}}
\makeatother

\usepackage[a4paper,text={160mm,247mm},centering]{geometry}

\allowdisplaybreaks[3]

\usepackage[font=small,labelfont=bf,width=0.85\textwidth]{caption}

\fi

\ifarxiv\else

\iftrue
\PassOptionsToClass{showpacs,showkeys,preprintnumbers}{revtex4-1}
\PassOptionsToClass{superscriptaddress}{revtex4-1}
\fi
\documentclass[10pt,letterpaper,twocolumn,amsmath,amssymb,final,aps,pra]{revtex4-1}

\usepackage{graphicx}
\usepackage[bookmarks=true,hyperfigures=true]{hyperref}
\usepackage[usenames, dvipsnames]{color}

\makeatletter
\let\o@a@f\@author@finish
\def\@author@finish{\o@a@f%
\let\@authors\empty\def\AF@opr##1{}\def\CO@opr##1##2##3{}%
\def\AU@opr##1##2##3{\ifx\@authors\@empty\toks@\expandafter{##2}%
  \else\toks@\expandafter{\@authors, ##2}\fi\edef\@authors{\the\toks@}}
\@AAC@list}
\makeatother

\fi

\expandafter\def\expandafter\bfseries\expandafter{\bfseries\ifmmode\else\boldmath\fi}
\expandafter\def\expandafter\mdseries\expandafter{\mdseries\ifmmode\else\unboldmath\fi}
\expandafter\def\expandafter\normalfont\expandafter{\normalfont\ifmmode\else\unboldmath\fi}

\usepackage{fixmath}

\RequirePackage[extdef]{delimset}

\let\barefrac=\frac
\renewcommand{\frac}[2]{\mathinner{\barefrac{#1}{#2}}}

\let\baresqrt=\sqrt
\makeatletter
\renewcommand{\sqrt}{\@ifnextchar[\@sqrt@space@a\@sqrt@space@b}
\def\@sqrt@space@a[#1]#2{\mathinner{\mathchoice{\mkern-3mu}{\mkern-3mu}{}{}\baresqrt[#1]{#2}}}
\def\@sqrt@space@b#1{\mathinner{\mathchoice{\mkern-3mu}{\mkern-3mu}{}{}\baresqrt{#1}}}
\makeatother

\makeatletter
\let\per@dot@old=\.
\def\.{\ifmmode\def\per@dot@sel{\mkern3mu}\else\def\per@dot@sel{\per@dot@old}\fi\per@dot@sel}
\makeatother

\newcommand{\sfrac}[2]{{\textstyle\mathord{\frac{#1}{#2}}}}

\newcommand{\vfrac}[2]{\ifmmode\mathinner{\textstyle^{#1}\!/\!_{#2}}\else$^{#1}\!/\!_{#2}$\fi}

\makeatletter\let\Re\@undefined\let\Im\@undefined\makeatother
\DeclareMathOperator{\Re}{Re}
\DeclareMathOperator{\Im}{Im}
\DeclareMathOperator{\tr}{tr}

\newcommand{\superN}{\mathcal{N}}

\def\[{\begin{equation}}
\def\]{\end{equation}}

\providecommand{\href}[2]{#2}

\makeatletter
\def\mr@ignsp#1 {\ifx\:#1\@empty\else #1\expandafter\mr@ignsp\fi}%
\newcommand{\multiref}[1]{\begingroup
\xdef\mr@no@sparg{\expandafter\mr@ignsp#1 \: }%
\def\mr@comma{}%
\@for\mr@refs:=\mr@no@sparg\do{\mr@comma\def\mr@comma{,}\ref{\mr@refs}}%
\endgroup}
\renewcommand{\eqref}[1]{(\multiref{#1})}
\makeatother

\makeatletter
\newcommand{\namedref}[2]{\hyperref[#2]{#1~\ref*{#2}}}
\newcommand{\secref}{\@ifstar{\namedref{Section}}{\namedref{Sec.}}}
\newcommand{\appref}{\@ifstar{\namedref{Appendix}}{\namedref{App.}}}
\newcommand{\tabref}{\@ifstar{\namedref{Table}}{\namedref{Tab.}}}
\newcommand{\figref}{\@ifstar{\namedref{Figure}}{\namedref{Fig.}}}
\makeatother

\let\oldbib=\thebibliography
\def\thebibliography{\phantomsection\addcontentsline{toc}{section}{\refname}\oldbib}

\let\oldtoc=\tableofcontents
\def\tableofcontents{\phantomsection\addcontentsline{toc}{section}{\contentsname}\oldtoc}

\providecommand{\hypersetup}[1]{}

\hypersetup{plainpages=false}
\hypersetup{pdfpagemode=UseNone}
\hypersetup{bookmarksnumbered=true}
\hypersetup{pdfstartview=FitH}
\hypersetup{colorlinks=false}
\hypersetup{citebordercolor={.5 1 .5}}
\hypersetup{urlbordercolor={.5 1 1}}
\hypersetup{linkbordercolor={1 .7 .7}}

\makeatletter
\let\@keywords\@empty
\let\@subject\@empty
\providecommand{\keywords}[1]{\gdef\@keywords{#1}}
\providecommand{\subject}[1]{\gdef\@subject{#1}}
\def\thetitle{\@title}
\def\theauthor{\@author}
\def\thesubject{\@subject}
\def\thedate{\@date}
\def\thekeywords{\@keywords}
\makeatother
\AtBeginDocument{
\hypersetup{pdftitle={\thetitle}}%
\hypersetup{pdfauthor={\theauthor}}%
\hypersetup{pdfsubject={\thesubject}}%
\hypersetup{pdfkeywords={\thekeywords}}%
}

\ifpublic
\newcommand{\remark}[2][]{\ignorespaces}
\else
\usepackage{ifpdf}
\ifpdf\else\RequirePackage[active]{srcltx}\fi
\RequirePackage{color}

\newcommand{\remark}[2][]{{\normalfont\sffamily\hspace{1ex}\def\tmparga{#1}%
  \def\tmpargb{MR}\ifx\tmparga\tmpargb\color{magenta}\fi%
  \def\tmpargb{NB}\ifx\tmparga\tmpargb\color{blue}\fi%
  \def\tmpargb{AG}\ifx\tmparga\tmpargb\color{Orange}\fi%
  \def\tmpargb{Refs}\ifx\tmparga\tmpargb\color{Orchid}\fi%
  \def\tmpargb{}\ifx\tmparga\tmpargb\color{red}\fi%
  \def\tmpargb{}\ifx\tmparga\tmpargb\else \textbf{#1:} \fi%
  #2\hspace{1ex}}}
\fi

\newcommand{\remarkref}[1][]{{\def\tmparga{#1}\def\tmpargb{}%
  \ifx\tmparga\tmpargb\remark[Refs]{needed}\else\remark[Refs]{#1}\fi}}

\begin{document}
\title{Untwisting the symmetries of $\beta$-deformed Super-Yang--Mills}
\long\def\theabstract{

We demonstrate that the planar real-$\beta$-deformed Super-Yang--Mills theory possesses an infinitely-dimensional Yangian symmetry algebra and thus is classically integrable. This is achieved by the introduction of the twisted coproduct which allows us to lift the apparent $\mathcal{N}=1$ supersymmetry first to the full $\mathcal{N}=4$ symmetry of the parent $\superN = 4$ SYM theory, and subsequently to its Yangian.
}

\ifarxiv\else

\author{Aleksander Garus}
\email{agarus@itp.phys.ethz.ch}
\affiliation{Institut f\"ur Theoretische Physik, 
Eidgen\"ossische Technische Hochschule Z\"urich, 
Wolfgang-Pauli-Strasse 27, 8093 Z\"urich, Switzerland}

\begin{abstract}
\theabstract
\end{abstract}

\maketitle
\fi

\ifarxiv
\pdfbookmark[1]{Title Page}{title}
\thispagestyle{empty}


\vspace*{2cm}
\begin{center}%
\begingroup\Large\bfseries\thetitle\par\endgroup
\vspace{1cm}

\begingroup\scshape
Aleksander Garus
\endgroup

\vspace{5mm}
\textit{Institut f\"ur Theoretische Physik,\\
Eidgen\"ossische Technische Hochschule Z\"urich,\\
Wolfgang-Pauli-Strasse 27, 8093 Z\"urich, Switzerland}
\vspace{0.1cm}

\begingroup\ttfamily\small
agarus@itp.phys.ethz.ch\par
\endgroup
\vspace{5mm}

\vfill

\textbf{Abstract}\vspace{5mm}

\begin{minipage}{12.7cm}
\theabstract
\end{minipage}

\vspace*{4cm}

\end{center}

\newpage
\fi

\section{Introduction}
\label{sec:intro}

The discovery of integrability (see \cite{Beisert:2010jr} for review) of the planar $\superN = 4$ Super-Yang--Mills (SYM) theory and its gravity dual, the SUGRA limit of type IIB superstring theory on $AdS_5 \times S^5$, allowed for what is up to date the most convincing tests of the AdS/CFT correspondence \cite{Maldacena:1997re}. Indeed, the huge amount of usually not obvious symmetries in those theories made computations of non-trivial observables on both sides feasible. The integrability techniques enabled to find some observables at any value of coupling, thus being an indisposable tool for testing the weak/strong duality.

After the first convincing results stemming from $\superN = 4$ SYM and ABJM theories (and their respective duals), a search for other examples of dualities began, ones that would allow to further verify Maldacena's proposal. 

A way of controllably obtaining such theories which proved very successful was to deform the initial ones in a way which preserves conformal invariance. For $\superN = 4$ SYM such deformations -- modifications of the superpotential -- were first discussed by Leigh and Strassler in \cite{Leigh:1995ep}. It was early observed that one of the deformations, the real-$\beta$-deformation, preserves integrability of $\superN = 4$ SYM (see \cite{Zoubos:2010kh} and references therein) in the planar limit. Its gravity dual was found by Lunin and Maldacena \cite{Lunin:2005jy}, and later demonstrated to stem from TsT transformations of the original $AdS_5 \times S^5$ background \cite{Frolov:2005ty}, \cite{Frolov:2005dj}. 

The real-$\beta$-deformed theory is manifestly an $\superN =1$ SYM. In \cite{Mansson:2008xv}, \cite{Dlamini:2016aaa} it was demonstrated however, that the manifest $SU(3) \times U(1)$ R-symmetry of $\superN = 4$ SYM expressed in the $\superN = 1$ language (see Section \ref{sec:betadef} ) does survive the deformation, albeit in a twisted way. Thus the $\superN = 4$ supersymmetry is not necessarily broken, but rather hidden. This result is backed up by the study of amplitudes in \cite{Khoze:2005nd}. The conclusion there was that the amplitudes in the twisted theory can be easily obtained from the ones of $\superN = 4$ SYM by a procedure that depends only on the \emph{external} legs, irrespectively of the internal structure, even though the vertices get deformed too.

The superconformal symmetry by itself is not sufficient to account for integrability of a field theory. For the parent $\superN = 4$ SYM the correct infinite-dimensional algebra has been identified as the Yangian $Y(\mathfrak{psu}(2,2|4))$ \cite{Dolan:2003uh}. It is natural to expect that the integrability of the real-$\beta$-deformed theory will be explained by a suitable deformation of this Yangian algebra. It was indeed shown in \cite{Ihry:2008gm} that some closed subsectors of it do enjoy a twisted Yangian symmetry, resonating well with the results of \cite{Beisert:2005if} on the twisted R-matrix.

In our recent work \cite{Beisert:2017pnr} we presented a novel framework which allows to establish a non-local symmetry of a given theory and used it to show the Yangian invariance of planar $\superN = 4$ SYM. In this work we will push our formalism further and demonstrate the extended symmetries of the real-$\beta$-deformed SYM, where nonlocality appears already at the level of R-symmetry.

The article is organized as follows. In section \ref{sec:betadef} we introduce the real-$\beta$-deformed SYM theory. That is followed by the construction of the twisted coproduct for R-symmetry generators so that we obtain the 12 missing supercharges of $\superN = 4$ SYM in section \ref{sec:untwist}. In section \ref{sec:yangian} we give a brief introduction to Yangian algebras, construct the twisted level-1 momentum generator $\widehat{P}_{\alpha \dot\alpha}$ and show that it is a symmetry of the theory. We comment on the results in section \ref{sec:conclusions}. The appendices contain all the formulae necessary to reproduce the results.

\section{$\beta$-deformed $\mathcal{N}=4$ SYM}
\label{sec:betadef}

The action of the real-$\beta$-deformed $\superN = 4$ SYM \cite{Leigh:1995ep} is most conveniently expressend in the $\superN = 1$ language, the field content being three chiral and one vector superfield. Working with component fields, $\Phi_i$ are the three complex scalar fields, $\lambda_{i \alpha}$ their superpartners ($i=1,\ 2,\ 3$). The gauge field $A_{\alpha \dot{\alpha}}$ has the gluino $\lambda_{4 \alpha}$ as its superpartner and acts as a connection for the covariant derivative $D_{\alpha \dot{\alpha}}=\partial_{\alpha \dot{\alpha}}+i A_{\alpha \dot{\alpha}}$. All the fields are in the adjoint of the gauge group $U(N)$. 

Written out explicitly, the Langrangian of the theory takes the form:

\begin{align}
\mathcal{L}=\tr \Bigg(& -\frac{1}{8}\epsilon^{\alpha \beta}\epsilon^{\gamma \kappa} \epsilon^{\dot{\alpha} \dot{\beta}} \epsilon^{\dot{\gamma}\dot{\kappa}} [D_{\alpha \dot{\beta}},D_{\gamma \dot{\kappa}}][D_{\beta \dot{\alpha}}, D_{\kappa \dot{\gamma}}]-\frac{1}{2}\epsilon^{\alpha \beta}\epsilon^{\dot{\alpha}\dot{\beta}}[D_{\alpha \dot{\alpha}},\bar{\Phi}^i][D_{\beta \dot{\beta}},\Phi_i]+ \nonumber \\
&-\frac{1}{2}[\Phi_i, \Phi_j]_{\beta_{ij}}[\bar{\Phi}^i,\bar{\Phi}^j]_{\beta_{ij}}+\frac{1}{4}[\Phi_i,\bar{\Phi}^i][\Phi_j,\bar{\Phi}^j]+\epsilon^{\alpha \beta} \epsilon^{\dot{\alpha} \dot{\beta}}\bar{\lambda}^4_{\dot{\beta}} [D_{\beta \dot{\alpha}},\lambda_{4 \alpha}]+\nonumber \\
&+ \epsilon^{\alpha \beta} \epsilon^{\dot{\alpha} \dot{\beta}}\bar{\lambda}^i_{\dot{\beta}} [D_{\beta \dot{\alpha}},\lambda_{i \alpha}]+i\left(\epsilon^{\alpha \beta}[\lambda_{4 \alpha},\lambda_{i \beta}]\bar{\Phi}^i+\epsilon^{\dot{\alpha} \dot{\beta}}[\bar{\lambda}^4_{\dot{\alpha}},\bar{\lambda}^i_{\dot{\beta}}]\Phi_i \right) + \nonumber \\
&+\frac{i}{2} \left(\epsilon^{ijk}\epsilon^{\alpha \beta} [\lambda_{i \alpha},\lambda_{j \beta}]_{\beta_{ij}}\Phi_k+ \epsilon_{ijk}\epsilon^{\dot{\alpha} \dot{\beta}}[\bar{\lambda}^i_{\dot{\alpha}},\bar{\lambda}^j_{\dot{\beta}}]_{\beta_{ij}}\bar{\Phi}^k \right)\Bigg) . 
\label{eq:n4sym_act_beta}
\end{align} 

In the above we have introduced a $\beta$-deformed graded commutator:
\begin{equation}
[f_i, g_j]_{\beta_{ij}}=e^{i \pi \beta_{ij}}f_i g_j - (-1)^{|f_i||g_j|}e^{i \pi \beta_{ji}} g_j f_i ,
\end{equation}
where $\beta_{ij}=-\beta_{ji}$, $\beta_{12}=\beta_{23}=\beta_{31}=\beta \in \mathbb{R}$.

The undotted and dotted spinor indices $\alpha$, $\dot{\alpha}$ take values $1,\ 2$. One can reintroduce the more common vector indices by contracting the spinor ones with Pauli matrices: $x_\mu = \frac{1}{ \sqrt{2} }\sigma_\mu^{\alpha \dot{\alpha}} x_{\alpha \dot{\alpha}}$.

Taking $\beta=0$ we recover the maximally symmetric $\mathcal{N}=4$ SYM theory. In the present form however, only $SU(3) \times U(1)$ subgroup of the $SU(4)$ R-symmetry is manifestly present, with the $U(1)$ factor corresponding to the vector superfield and the $SU(3)$ to 3 chirals. 

For an aribtrary real $\beta$ the action \eqref{eq:n4sym_act_beta} is invariant under $\superN = 1$ supersymmetry with the charges $Q_{\alpha}$, $\bar{Q}_{\dot{\alpha}}$, which in the undeformed theory correspond to $Q^4_{\alpha}$, $\bar{Q}_{4 \dot{\alpha}}$ (see Appendix B).

As was shown in \cite{Leigh:1995ep}, the theory remains conformal even on a quantum level -- a property shared with the full $\superN =4$ SYM -- and thus yields another example of a superconformal quantum field theory, as already alluded to in the Introduction. 

\section{Untwisting the twist}
\label{sec:untwist}
As already discussed in the Introduction, the $\beta$-deformed theory shares a lot of similarities with its parent, the $\superN = 4$ SYM. In this section we want to show that the deformed theory in fact possesses the full (albeit deformed and non-local) $\superN = 4$ supersymmetry.
To this end it is enough to demonstrate that the theory is invariant under the full $SU(4)$ R-symmetry group. All the remaining supersymmetry generators can be then recovered from the $Q_\alpha=Q^4_{\alpha}$ and $\bar{Q}_{\dot\alpha}=\bar{Q}_{4 \dot\alpha}$ via:

\[
Q^i_{\alpha}=[R^i {}_4, Q_\alpha]
\label{eq:miss_sca}
\]
and similarly for the conjugate ones.

Corresponding to the $\superN=1$ supersymmetry, the action is invariant under the $R^4 {}_4$ component of the R-symmetry (in this case the $U(1)$ generator). In  the case of a planar limit of the real-$\beta$ deformation however, the symmetry is actually larger, as all the diagonal elements $R^c {}_c$ survive. The additional $U(1)$ symmetries enabled the construction of the background of the gravity dual \cite{Lunin:2005jy} (see also \cite{Mansson:2008xv} and \cite{Dlamini:2016aaa} for a discussion of extended symmetry of the theory).

We will now show that all the remaining generators $R^a {}_b$ (see Appendix A for the detailed action) can be promoted to the symmetries of the action \eqref{eq:n4sym_act_beta}.
This will be achieved by a suitable modification of the coproduct of the generators, i.e. the action on products of fields.
In the real-$\beta$-deformed SYM this is achieved by the Drinfeld-Reshetikhin twist of the comultiplication \cite{vanTongeren:2015soa}:

\begin{equation}
\Delta_\mathcal{F} = \mathcal{F} \Delta \mathcal{F}^{-1}, 
\end{equation}  

\noindent with $\mathcal{F}$ given in terms of $SU(4)$ charges of the fields:

\begin{equation}
\mathcal{F}=e^{i \pi \beta (h_1 \wedge h_2 + h_2 \wedge h_3 - h_1 \wedge h_3)},
\end{equation}
where $h_i = R^i {}_{i}+R^4 {}_4$ (no sum over $i$).

We will omit the subscript $\mathcal{F}$ from now on.
As alluded to above, for the diagonal R-symmetry generators the modification acts as identity:

\begin{equation}
\Delta R^c {}_c = \mathbb{I} \otimes R^c {}_c + R^c {}_c \otimes \mathbb{I} .
\label{eq:nondef_copr}
\end{equation}

\noindent The coproduct of non-diagonal elements however changes to

\begin{equation}
\Delta R^a {}_b = \mathbb{K}_{ab} \otimes R^a {}_b + R^a {}_b \otimes \mathbb{K}_{ba}
\label{eq:def_copr}
\end{equation}
where $\mathbb{K}_{ba}=\mathbb{K}_{ab}^{-1}$. The element $\mathbb{K}_{ab}$ is group-like:

\begin{equation}
\Delta \mathbb{K}_{ab} = \mathbb{K}_{ab} \otimes \mathbb{K}_{ab}. 
\label{eq:kr_cop}
\end{equation}

\noindent Its action on the fields of the theory amounts to a multiplication by a $\beta$-dependent phase (see Appendix A).
Observe that the coproduct \eqref{eq:kr_cop} does not cope well with cyclicity. It is a usual problem with non-trivial coproducts, since due to appearance of trace in \eqref{eq:n4sym_act_beta} (and any other Lagrangian) the action of a theory is cyclic. To circumvent this problem, in \cite{Beisert:2017pnr} we developed an equation-of-motion-based formalism which is equivalent to invariance of action. We will sketch it here. For a usual Lie-type symmetry with a trivial coproduct which leaves the action invariant:

\[
J S := (J Z_a) \frac{\delta S}{\delta Z_a} = 0,
\label{eq:jseq0}
\]
where $J$ is an algebra generator, we can differentiate \eqref{eq:jseq0} with respect to an arbitrary field $Z_c$ to obtain:

\[
J \frac{\delta S}{\delta Z_c} = - \frac{\delta (J Z_a)}{\delta Z_c} \frac{\delta S}{\delta Z_a}.
\label{eq:magic_lvl0}
\]

Observe that the equality \eqref{eq:magic_lvl0}, contrary to \eqref{eq:jseq0}, contains no cyclic objects. What \eqref{eq:magic_lvl0} states is that the result of $J$ acting on an equation of motion $\frac{\delta S}{\delta Z_c}$ is a particular combination of other equations of motion with coefficients determined purely by a representation of $J$ on fields $Z_a$ of the theory. We argued in \cite{Beisert:2017pnr} that \eqref{eq:magic_lvl0} is equivalent to the invariance of the action. 
The question now is what is the equivalent of \eqref{eq:magic_lvl0} in case of a twisted coproduct \eqref{eq:kr_cop}. 
Surprisingly, the answer is that \eqref{eq:magic_lvl0} holds in this case in an unchanged form, with no explicit appearance of generators $\mathbb{K}_{ab}$ (even though they do contribute while acting on an equation of motion) - a fact we confirmed by direct computation. 
Indeed, it can actually be shown that their action amounts to an overall factor while acting on equations of motion and hence becomes unobservable. Heuristically this can be explained by noting that the kinetic term is always undeformed. All the transformations are thus fully determined by the single-field action of $R^a {}_b$. Having shown that all the R-symmetry generators are symmetries of the action, we now may use them to construct the missing supercharges according to \eqref{eq:miss_sca} and thus argue that indeed the real-$\beta$-deformed SYM possesses a $\superN = 4$ supersymmetry. The formula \eqref{eq:magic_lvl0} holds then also for the generators $Q^a_{\alpha}$ and $\bar{Q}_{a \dot\alpha}$, as we verified again by explicit calculation.

Concluding this paragraph, let us mention that it may actually be directly shown that the action obtained by integrating the Lagrangian \eqref{eq:n4sym_act_beta} is invariant under all R-symmetry generators $R^a {}_b$, without resorting to the equations of motion. To this end, we cannot rely on picking an arbitrary cyclic representative, but rather need to consider the action as the averaged sum of all the possible ones . This of course is a trivial operation under the trace, but in order to act with a generator of transformation, we need to \emph{cut the trace open} at some point. A trivial coproduct respects cyclicity of the trace and hence the choice of the opening point is irrelevant, but that ceases to be true for coproducts like \eqref{eq:def_copr}. We hence rewrite the terms forming the action as:

\[
\text{tr} \left( Z_1 Z_2 ... Z_n \right) \rightarrow  \frac{1}{n} \sum_{\sigma \in \mathbb{Z}_n} Z_{\sigma(1)} Z_{\sigma(2)} ... Z_{\sigma(n)},
\] 
and then act on them using iterated coproducts. A heuristic picture to have in mind is cutting open the closed (due to trace) chain of fields at every possible point. Under so defined application of the generators, the action is invariant.
As this approach however does not easily generalise to higher level Yangian symmetries discussed in the next section, we will not discuss it here in more detail.


\section{Twisted Yangian}
\label{sec:yangian}

Having established a twisted $\superN = 4$ superconformal symmetry of the real-$\beta$-deformed theory, we are now in position to observe what happens to the Yangian symmetry. Indeed, for the $\superN = 4$ SYM the Yangian of $\mathfrak{psu} (2,2|4)$ underlies its integrability. Twisted Yangian algebra has been identified in \cite{Ihry:2008gm} as the symmetry of some subsectors of the real-$\beta$-deformed SYM. We would now like to promote this discussion to a full theory, taking into account the nonlinearities of the symmetries. 

The Yangian algebra is built on a Lie (super)-algebra as follows \cite{Drinfeld:1985rx}, \cite{Loebbert:2016cdm}. Starting from the usual commutation relations:

\[
[J^A,J^B \} = f^{AB} {}_C J^C,
\]
we introduce a new set of \emph{level-1} generators $\widehat{J}^A$ transforming in the adjoint of the original algebra:

\[
[J^A, \widehat{J}^B \} = f^{AB} {}_C \widehat{J}^C.
\]

An infinite tower of higher-level generators is obtained by the commutation of the lower level ones, i.e. level-2 generators can be obtained by commuting two level-1 ones. The generators are subject to Serre relations, which we will not discuss in the current work.

The crucial feature of the level-1 generators is their non-trivial coproduct:
\[
\Delta \widehat{J}^A = \mathbb{I} \otimes \widehat{J}^A + \widehat{J}^A \otimes \mathbb{I} + f^A {}_{BC} J^B \otimes J^C,
\]
which features two level-0 generators acting simultaneously. This coproduct renders the application of level-1 Yangian generators directly to the action cumbersome, which is why we work with the equations of motion of the theory (see \cite{Beisert:2017pnr} for details).

In the case of $\mathcal{N}=4$ SYM, the simplest level-1 Yangian generator is the level-1 momentum $\widehat{P}_{\alpha \dot{\alpha}}$, whose coproduct is given by:

\begin{align}
\Delta \widehat{P}_{\alpha \dot{\alpha}, \ \superN=4} &= \mathbb{I}\otimes \widehat{P}_{\alpha \dot\alpha} + \widehat{P}_{\alpha \dot\alpha} \otimes \mathbb{I} + D \wedge P_{\alpha \dot\alpha} + P_{\beta \dot\alpha} \wedge L^\beta {}_\alpha  + P_{\alpha \dot\beta} \wedge \bar{L}^{\dot\beta} {}_{\dot\alpha} \nonumber \\
& -\sfrac{1}{2} Q_{\alpha} \wedge \bar{Q}_{\dot\alpha} -\sfrac{1}{2} Q^i_{\alpha} \wedge \bar{Q}_{i \dot\alpha} \nonumber \\
&=\mathbb{I}\otimes \widehat{P}_{\alpha \dot\alpha} + \widehat{P}_{\alpha \dot\alpha} \otimes \mathbb{I} + h^A {}_{BC} J^B \otimes J^C,
\end{align}

\noindent where $h^A {}_{BC}$ are the $\mathfrak{psu}(2,2|4)$ structure constants.
Since the generators of conformal algebra and $Q_{\alpha}$ and $\bar{Q}_{\dot{\alpha}}$ are insensitive to the $\beta$-deformation, the only possible twists in the coproduct for $\widehat{P}_{\alpha \dot{\alpha}}$ can appear in the terms containing $Q^i_{\alpha}$ and $\bar{Q}_{i \dot{\alpha}}$. Indeed, following the methods developed in \cite{Beisert:2017pnr} we checked via explicit computations that the modified coproduct:

\begin{align}
\Delta \widehat{P}_{\alpha \dot{\alpha}, \ \beta \neq 0} &= \mathbb{I}\otimes \widehat{P}_{\alpha \dot\alpha} + \widehat{P}_{\alpha \dot\alpha} \otimes \mathbb{I} + D \wedge P_{\alpha \dot\alpha} + P_{\beta \dot\alpha} \wedge L^\beta {}_\alpha  + P_{\alpha \dot\beta} \wedge \bar{L}^{\dot\beta} {}_{\dot\alpha} \nonumber \\
& -\sfrac{1}{2} Q_{\alpha} \wedge \bar{Q}_{\dot\alpha} -\sfrac{1}{2} Q^i_{\alpha} \mathbb{K}_{i \alpha}^{-1} \otimes \bar{Q}_{i \dot\alpha} \bar{\mathbb{K}}_{i \dot\alpha} - \sfrac{1}{2} \bar{Q}_{i \dot\alpha}\bar{\mathbb{K}}_{i \dot\alpha}^{-1} \otimes Q^i_{\alpha} \mathbb{K}_{i \alpha} 
\label{eq:phat_def_copr}
\end{align}

\noindent maps equations of motion of the theory to each other, provided the single-field action of $\widehat{P}_{\alpha \dot\alpha}$ is given by eq. \eqref{eq:phat_single}

\begin{align}
&\widehat{P}_{\alpha \dot\alpha} \Phi_i = 0 \nonumber \\
&\widehat{P}_{\alpha \dot\alpha} \bar{\Phi}^i = 0 \nonumber \\
&\widehat{P}_{\alpha \dot\alpha} \lambda_{i \beta} = i \epsilon_{\alpha \beta} \epsilon_{ijk} \{\bar{\lambda}^j_{\dot\alpha}, \bar{\Phi}^k \}_{\beta_{jk}} -i \epsilon_{\alpha \beta} \{ \Phi_i, \bar{\lambda}_{4 \dot\alpha} \} \nonumber \\
&\widehat{P}_{\alpha \dot\alpha} \bar{\lambda}^i_{\dot\beta} = i \epsilon_{\dot\alpha \dot\beta} \epsilon^{ijk} \{\lambda_{j \alpha}, \Phi_k \}_{\beta_{jk}} -i \epsilon_{\dot\alpha \dot\beta} \{ \bar{\Phi}^i, \lambda_{4 \alpha} \} \nonumber \\
&\widehat{P}_{\alpha \dot\alpha} \lambda_{4 \beta} = i \epsilon_{\alpha \beta}  \{\bar{\lambda}^i_{\dot\alpha}, \Phi_i \} \nonumber \\
&\widehat{P}_{\alpha \dot\alpha} \bar{\lambda}_{4 \dot\beta} = i \epsilon_{\dot\alpha \dot\beta}  \{\lambda_{i \alpha}, \bar{\Phi}^i \} \nonumber \\
&\widehat{P}_{\alpha \dot\alpha} D_{\beta \dot\beta} = -\epsilon_{\alpha \beta} \epsilon_{\dot\alpha \dot\beta} \{\Phi_i, \bar{\Phi}^i \}.
\label{eq:phat_single}
\end{align}

 Observe that the modified part is no longer antisymmetric. Furthermore, due to the group-like nature of twist generators $\mathbb{K}$ the coproduct is no-longer \emph{bilocal}, but actually may encompass all the fields from the product. This is in contrast with the undeformed theory, but already made its appearance at level-0.

In the case of undeformed Yangian algebras we showed that the following formula, an level-1 analogue of \eqref{eq:magic_lvl0}, is equivalent to the invariance of the action and utilises only equations of motion, thus circumventing the issue of the cyclicity:

\[
\widehat{J}^A \frac{\delta S}{\delta Z_m} = -\frac{\widehat{J}^A Z_n}{\delta Z_m} \frac{\delta S}{\delta Z_n} + h^A {}_{BC} \frac{\delta S}{\delta Z_n } \left(J^B \wedge \frac{\delta}{\delta Z_m} \right) (J^C Z_n).
\]

In case the of real-$\beta$-deformed theory, this equality gets a predictable deformation for the $\widehat{P}_{\alpha \dot\alpha}$ generator:

\begin{align}
&\widehat{P}_{\alpha \dot\alpha} \frac{\delta S}{\delta Z_m} = -\frac{\widehat{P}_{\alpha \dot\alpha} Z_n}{\delta Z_m} \frac{\delta S}{\delta Z_n} + f^{P_{\alpha \dot\alpha}} {}_{BC} \frac{\delta S}{\delta Z_n } \left(J^B \wedge \frac{\delta}{\delta Z_m} \right) (J^C Z_n)+ \nonumber \\ 
&\frac{\delta S}{\delta Z_n} \left(Q^i_{\alpha} \mathbb{K}_{i \alpha}^{-1} \wedge \frac{\delta }{\delta Z_m} \right) (\bar{Q}_{i \dot\alpha} \bar{\mathbb{K}}_{i \dot\alpha} \ Z_n) + \frac{\delta S}{\delta Z_n} \left(\bar{Q}_{i \dot\alpha}\bar{\mathbb{K}}_{i \dot\alpha}^{-1} \wedge \frac{\delta }{\delta Z_m}\right)(Q^i_{\alpha} \mathbb{K}_{i \alpha} \  Z_n) ,
\label{eq:magic_def}
\end{align}

\noindent where $f^A {}_{BC}$ are $\mathfrak{psu}(2,2|1)$ structure constants. Indeed, we checked that the equations of motion of the real-$\beta$-deformed SYM satisfy \eqref{eq:magic_def}.

Eventually, we see that the properly defined level-1 Yangian generator $\widehat{P}_{\alpha \dot\alpha}$ is a symmetry of real-$\beta$-deformed $\superN =4$ SYM. We thus obtain an infinitely-dimensional symmetry algebra also for the case of the twisted theory, which explains its observed integrability. On the level of algebra, the symmetry is just the Yangian $Y\left(\mathfrak{psu}(2,2|4) \right)$, the difference with the $\superN = 4$ SYM being only the Hopf structure.

\section{Comments and conclusions}
\label{sec:conclusions}

In this work we demonstrated that the planar real-$\beta$-deformed SYM theory \eqref{eq:n4sym_act_beta} possesses a much richer set of symmetries than manifest on the level of the action. By the use of a twisted coproduct we applied all the $\superN = 4$ SUSY generators and verified that they leave the action invariant and thus the planar real-$\beta$-deformed SYM shares the symmetry algebra $\mathfrak{psu}(2,2|4)$ with the $\superN = 4$ SYM. We then showed that those results can be lifted to the full Yangian $Y\left(\mathfrak{psu}(2,2|4) \right)$ symmetry of the theory.

Our result not only generalizes earlier constructions of Yangian algebra in the case of a deformed theory \cite{Ihry:2008gm}, but moreover verifies that the formalism we introduced in \cite{Beisert:2017pnr} can also be used to study deformed theories, with the main result, \eqref{eq:magic_def} being in a clear correspondence with the coproduct \eqref{eq:phat_def_copr} of the generator under consideration.
Our formalism can thus serve as a classical integrability check for a wide class of theories.

The above results are clearly purely classical. Though very close on the classical level, the undeformed and deformed theory differ when loop effects are considered (i.e. wrapping corrections appear already at 1-loop for the deformed theory, whereas they are postponed to higher loops for the $\superN = 4$ SYM). It would be interesting to see, what is the fate of Yangian symmetry once quantum corrections are introduced and whether it differs from the behavior of the parent theory. Even though the $\superN = 1$ superconformal symmetry persists, it is not known what happens to twisted generators, and thus also to higher level Yangian ones.

\pdfbookmark[1]{Acknowledgements}{ack}
\section*{Acknowledgements}

I would like to thank N. Beisert, R. Hecht, B. Hoare, M. Rosso and M. Sprenger for valuable discussions and comments.
The work is partially supported by grant no.\ 615203 
from the European Research Council under the FP7
and by the Swiss National Science Foundation
through the NCCR SwissMAP.


\ifarxiv
\bibliographystyle{nb}
\fi
\bibliography{BetaDef} 


\section*{Appendix A: Action of twisted R-symmetry}
\label{sec:AppA}

All R-symmetry generators $R^a {}_b$ annihilate the gauge field $A_{\alpha \dot\alpha}$. On the fermions they act as:

\[
R^a {}_b \lambda_{c \alpha} = \delta^a_c \lambda_{b \alpha} - \frac{1}{4} \delta^a_b \lambda_{c \alpha}.
\]
Action on the scalars is obtained by writing the complex fields $\Phi_i$ in terms of hermitian scalars with antisymmetric $\mathfrak{su}(4)$ indices: $\Phi_i \propto \phi_{i4}$. Then the R-symmetry generators act on them just like on the product of two fermions: $\phi_{i4} \sim \lambda_i \lambda_4 $. This also leads to fields $\Phi_i$ and $\bar{\Phi}^i$ mixing under the action of R-symmetry.

As already presented in \eqref{eq:def_copr}, the coproduct for $R^a {}_b$ receives a twist $\mathbb{K}_{ab}$. The group-like element $\mathbb{K}_{ab}$ acts on fields in the following way:

\begin{align}
\mathbb{K}_{ab} \Phi_i &= e^{i \pi r(a,b,i) \beta} \Phi_i \nonumber \\
\mathbb{K}_{ab} \bar{\Phi}^i &= e^{- i \pi r (a,b,i) \beta} \bar{\Phi}^i \nonumber \\
\mathbb{K}_{ab} \lambda_{i \alpha} &= e^{i \pi r(a,b,i) \beta} \lambda_{i \alpha} \nonumber \\
\mathbb{K}_{ab} \bar{\lambda}^i_{\alpha} &= e^{- i \pi r (a,b,i) \beta} \bar{\lambda}^i_{\alpha} \nonumber \\
\mathbb{K}_{ab} Z &= Z,
\end{align}
where $Z$ stands for all the remaining fields. The function $r(a,b,i)$ is given by:

\[
r(a,b,i)=(\delta^a_4 + \delta^4_b) \sum_{c=1}^4 \left( \varepsilon_{abic}+(1- \delta^a_4 - \delta^4_b) (-2 \epsilon_{abc}+(1-|\epsilon_{abc}|)sig(a,b)) \right).
\]
The antisymmetric function $sig(a,b)=-sig(b,a)$ is given by: $sig(1,2)=sig(2,3)=sig(3,1)=1$, $sig(a,4)=0$. 


\section*{Appendix B: Action of supersymmetry generators}
\label{sec:AppB}

\subsection*{Non-deformed $\superN = 1$ supersymmetry}

The manifest $\superN = 1$ SUSY generators $Q_{\alpha}$ and $\bar{Q}_{\dot\alpha}$ act on the fields in the following way:

\begin{align}
Q_\alpha \Phi_i &  = \sqrt{2} i \lambda_{i \alpha} \\
Q_\alpha \bar{\Phi}^i & = 0 \\
Q_\alpha  \lambda_{i \beta} & =-\frac{1}{\sqrt{2}} \epsilon_{\alpha \beta} \epsilon_{ijk} [\bar{\phi}^j,\bar{\phi}^k]_{\beta_{jk}} \\
Q_\alpha  \bar{\lambda}^i_{\dot{\beta}} & = -\sqrt{2}i \sigma^\mu_{\alpha \dot{\beta}} [D_\mu, \bar{\phi}^i]\\
Q_\alpha  \lambda_{4 \beta} &= -\frac{1}{\sqrt{2}} \epsilon^{\dot{\gamma} \dot{\kappa}} [D_{\alpha \dot{\gamma}}, D_{\beta \dot{\kappa}}] -\frac{1}{\sqrt{2}} \epsilon_{\alpha \beta} [\phi_i, \bar{\phi}^i] \\
Q_\alpha  \bar{\lambda}^4_{\dot{\beta}} &=0 \nonumber \\
Q_\alpha  D_{\beta \dot{\gamma}} & = -\sqrt{2} \epsilon_{\alpha \beta} \bar{\lambda}^4_{\dot{\gamma}} ,
\end{align}

\noindent and analogous formulae for $\bar{Q}_{\dot \alpha}$.

\subsection*{The hidden supersymmetry generators}

The action of $Q^i_{ \alpha}$ and $\bar{Q}_{i \dot\alpha}$ is to be obtained by commuting the manifest supersymmetry generators with R-symmetry generators: $Q^i_{\alpha}=[R^i {}_4, Q_\alpha]$. For the sake of completeness, since they appear explicitly in the deformed coproduct for the level-1 momentum generator \eqref{eq:phat_def_copr}, we present the results explicitly.

\begin{align}
&Q^i_{\alpha} \Phi_j = \sqrt{2} i \delta^i_j \lambda_{4 \alpha} \nonumber \\ 
&Q^i_{\alpha} \bar{\Phi}^j = \sqrt{2} i \epsilon^{ijk} \lambda_{k \alpha} \nonumber \\
&Q^i_{\alpha} \lambda_{j \beta} =  \frac{1}{\sqrt{2}}\delta^i_j \left( \epsilon^{\dot\alpha \dot\beta}[D_{\alpha \dot\alpha},D_{\beta \dot\beta}] + \epsilon_{\alpha \beta} [\Phi_k,\bar{\Phi}^k] \right)-\sqrt{2}\epsilon_{\alpha \beta}[\Phi_j,\bar{\Phi}^i]_{\beta_{ij}} \nonumber \\
&Q^i_{\alpha} \bar{\lambda}^j_{\dot \alpha} = \sqrt{2} i \epsilon^{ijk} [D_{\alpha \dot\alpha}, \Phi_k]\nonumber \\
&Q^i_{\alpha} \lambda_{4 \beta}=- \epsilon_{\alpha \beta}\sqrt{2}\epsilon^{ijk} [\Phi_j, \Phi_k]_{\beta_{jk}} \nonumber \\
&Q^i_{\alpha} \bar{\lambda}^4_{\dot\alpha} = -\sqrt{2} [D_{\alpha \dot{\alpha}},\bar{\Phi}^i] \nonumber \\
&Q^i_{\alpha} D_{\beta \dot\beta}=\sqrt{2} \epsilon_{\alpha \beta} \bar{\Psi}^i_{\dot\beta}
\label{eq:qi_sf_act}
\end{align}

\noindent Again similar formulae hold for $\bar{Q}_{i \dot\alpha}$.

The coproduct for the generators $Q^i_{\alpha}$ gets twisted:

\[
\Delta Q^i_{\alpha} = \mathbb{K}_{i \alpha} \otimes Q^i_{\alpha} + Q^i_\alpha \otimes \mathbb{K}^{-1}_{i \alpha}.
\]

The twist generator $\mathbb{K}_{i \alpha}$ has the following action on fields:

\begin{align}
&\mathbb{K}_{i \alpha} \Phi_j = e^{i \pi sig(i,j) \beta} \Phi_j \nonumber \\
&\mathbb{K}_{i \alpha} \bar{\Phi}^j = e^{-i \pi sig(i,j) \beta} \bar{\Phi}^j \nonumber \\
&\mathbb{K}_{i \alpha} \lambda_{j\gamma} = e^{i \pi sig(i,j) \beta} \lambda_{j \gamma} \nonumber \\
&\mathbb{K}_{i \alpha} \bar{\lambda}^j_{\dot\alpha} = e^{-i \pi sig(i,j) \beta} \bar{\lambda}^j_{\dot\alpha} \nonumber \\
&\mathbb{K}_{i \alpha} Z = Z ,
\end{align}
where the $sig(a,b)$ function has been defined in Appendix A.


\section*{Appendix C: Conformal symmetry generators}
\label{sec:AppC}

In order to introduce the level-1 momentum generator and its coproduct \eqref{eq:phat_def_copr} we needed the generators of the conformal algebra: the dilatation $D$, the momentum $P_{\alpha \dot\alpha}$ and the Lorentz rotations $L^\alpha{}_\beta$ and $\bar{L}^{\dot\alpha} {}_{\dot\beta}$. Their explicit actions on the fields are:

\begin{align}
&P_{\alpha \dot\alpha} Z_m = [D_{\alpha \dot\alpha}, Z_m] \nonumber \\
&D Z_m = -x^{\alpha \dot\alpha} [D_{\alpha \dot\alpha}, Z_m] -\Delta_{Z_m} Z_m,
\end{align} 
where $\Delta_{Z_m}$ is a conformal weight of the field, equal to $1$ for scalars, $\frac{3}{2}$ for fermions and $0$ for the covariant derivative.

The Lorentz generators act in a following  way:

\begin{align}
L^\alpha {}_\beta \Phi_i &= -x^{\alpha \dot\alpha} [D_{\beta \dot\alpha}, \Phi_i]+\sfrac{1}{2} \delta^\alpha_\beta x^{\kappa \dot\kappa} [D_{\kappa \dot\kappa}, \Phi_i] \nonumber \\
L^\alpha {}_\beta \bar{\Phi}^i &= -x^{\alpha \dot\alpha} [D_{\beta \dot\alpha}, \bar{\Phi}^i]+\sfrac{1}{2} \delta^\alpha_\beta x^{\kappa \dot\kappa} [D_{\kappa \dot\kappa}, \bar{\Phi}^i] \nonumber \\
L^\alpha {}_\beta \lambda_{a \gamma} & = -x^{\alpha \dot\alpha} [D_{\beta \dot\alpha}, \lambda_{i \alpha}]+\sfrac{1}{2} \delta^\alpha_\beta x^{\kappa \dot\kappa} [D_{\kappa \dot\kappa}, \lambda_{a \alpha}]-\delta^\alpha_{\gamma} \lambda_{a \beta} + \delta^\alpha_\beta \lambda_{a \gamma} \nonumber \\
L^\alpha {}_\beta \bar{\lambda}^a_{\dot\gamma} & =-x^{\alpha \dot\alpha} [D_{\beta \dot\alpha}, \bar{\lambda}^a_{\dot\gamma}]+\sfrac{1}{2} \delta^\alpha_\beta x^{\kappa \dot\kappa} [D_{\kappa \dot\kappa}, \bar{\lambda}^a_{\dot\gamma}] \nonumber \\
L^\alpha {}_\beta D_{\gamma \dot\gamma} & =-x^{\alpha \dot\alpha} [D_{\beta \dot\alpha}, D_{\gamma \dot\gamma}]+\sfrac{1}{2} \delta^\alpha_\beta x^{\kappa \dot\kappa} [D_{\kappa \dot\kappa}, D_{\gamma \dot\gamma}] ,
\end{align}
with analogous expressions for conjugate generators $\bar{L}^{\dot\alpha} {}_{\dot\beta}$.


\section*{Appendix D: Equations of motion}
\label{sec:AppD}

Since we work with equations of motion in sections \ref{sec:untwist} and \ref{sec:yangian}, we list the variations of the action here:
\begin{align}
\frac{\delta S}{\delta D_{\gamma \dot{\kappa}}}&=-\frac{1}{2}\epsilon^{\alpha \beta}\epsilon^{\gamma \kappa} \epsilon^{\dot{\alpha}\dot{\beta}}\epsilon^{\dot{\gamma}\dot{\kappa}}[D_{\alpha \dot{\beta}},[D_{\beta \dot{\alpha}},D_{\kappa \dot{\gamma}}]]-\frac{1}{2}\epsilon^{\alpha \gamma}\epsilon^{\dot{\alpha}\dot{\kappa}}\left([\bar{\Phi}^i,[D_{\alpha \dot{\alpha}},\Phi_i]]+[\Phi_i,[D_{\alpha \dot{\alpha}},\bar{\Phi}^i]] \right)\nonumber \\
&+\epsilon^{\alpha \gamma}\epsilon^{\dot{\kappa} \dot{\beta}}\{\bar{\lambda}^4_{\dot{\beta}},\lambda_{4\alpha}\}+\epsilon^{\alpha \gamma}\epsilon^{\dot{\kappa}\dot{\beta}}\{\bar{\lambda}^i_{\dot{\beta}},\lambda_{i \alpha}\}\\
\frac{\delta S}{\delta \bar{\Phi}^i}&=\frac{1}{2}\epsilon^{\alpha \beta}\epsilon^{\dot{\alpha}\dot{\beta}}[D_{\alpha \dot{\alpha}},[D_{\beta \dot{\beta}},\Phi_i]]-\frac{1}{2}[\bar{\Phi}^j,[\Phi_i,\Phi_j]_{\beta_{ij}}]_{\beta_{ij}}+\frac{1}{4}[\Phi_i,[\Phi_j,\bar{\Phi}^j]] \nonumber \\
&+i\epsilon^{\alpha \beta} \{\lambda_{4 \alpha},\lambda_{i \beta} \}+\frac{i}{2}\epsilon_{ijk}\epsilon^{\dot{\alpha}\dot{\beta}} \{\bar{\lambda}^j_{\dot{\alpha}},\bar{\lambda}^k_{\dot{\beta}} \}_{\beta_{jk}} \\
\frac{\delta S}{\delta \bar{\lambda}^4_{\dot{\alpha}}}&=\epsilon^{\alpha \beta} \epsilon^{\dot{\beta}\dot{\alpha}}[D_{\beta \dot{\beta}},\lambda_{4 \alpha}]+i\epsilon^{\dot{\alpha}\dot{\beta}}[\bar{\lambda}^i_{\dot{\beta}},\Phi_i]\\
\frac{\delta S}{\delta \bar{\lambda}^i_{\dot{\alpha}}}&=\epsilon^{\alpha \beta} \epsilon^{\dot{\beta}\dot{\alpha}}[D_{\beta \dot{\beta}},\lambda_{i \alpha}]-i \epsilon^{\dot{\alpha}\dot{\beta}}[\bar{\lambda}^4,\Phi_i]+\frac{i}{2}\epsilon_{ijk}\epsilon^{\dot{\alpha}\dot{\beta}}[\bar{\lambda}^j_{\dot{\beta}},\bar{\Phi}^k]_{\beta_{ij}}
\end{align}
together with their respective conjugates. Putting fields on-shell corresponds of course to setting $\frac{\delta S}{\delta Z_m}=0$, where $Z_m$ stands for any of the fields.

\end{document}